\documentclass[referee,useAMS,usenatbib]{mn2e}

\usepackage{graphicx}

\title[QPO in Frequency in 1998 outburst of XTE J1550-564]
{Accretion flow behaviour during the evolution of the Quasi Periodic
Oscillation Frequency of XTE J1550-564 in 1998 outburst}
\author[Sandip K. Chakrabarti, Broja G. Dutta,  P.S. Pal]
{ Sandip K. Chakrabarti$^{1,2}$,
Broja G. Dutta$^{2,3}$, P.S. Pal$^{2}$\\
$^{1}$S. N. Bose National Centre for Basic Sciences, Salt Lake,
              Kolkata 700098, India\\
$^{2}$Indian Centre for Space Physics, Chalantika 43, Garia Station Rd., 
	     Kolkata, 700084, India\\
$^{3}$Y.S. Palapara College, Palpara, Purba-Medinipur, 721458, India}
\begin{document}

\date{}


\maketitle

\label{firstpage}

\begin{abstract}
Low and Intermediate Frequency Quasi-Periodic Oscillations (QPOs) 
are thought to be due to oscillations of Comptonizing regions or hot 
regions embedded in Keplerian discs. Observational evidence of 
evolutions of QPOs would therefore be very important as they throw 
lights on the dynamics of the hotter region.
Our aim is to find systems in which there is a well defined correlation
among the frequencies of the Quasi Periodic Oscillations over a range of
time so as to understand the physical picture.
In this paper, we concentrate on the archival data of XTE J1550-564 obtained
during 1998 outburst, and study the systematic drifts during the rising
phase from the $7^{th}$ of September 1998 to the $19^{th}$ September 
1998, when the QPO frequency increased monotonically from 
$81$mHz to $13.1$Hz. Immediately after that QPO frequency started to decrease 
and on the $26^{th}$ of September 1998, the QPO frequency 
became $2.62$Hz. After that its value remained almost constant.
This frequency drift can be modeled satisfactorily with a propagatory 
oscillating shock solution where the post-shock region behaves
as the Comptonized region. Comparing with the nature of a more recent 2005 
outburst of another black hole candidate GRO 1655-40, where QPOs disappeared 
at the end of the rising phase, we conjecture that this so-called 
'outburst' may not be a full-fledged outburst. 
\end{abstract}

\begin{keywords}
{accretion, accretion disc - shock waves - stars: individual (XTE J1550-564).}
\end{keywords}

\section{Introduction}

The Galactic soft X-ray transient (SXT) and a black hole candidate XTE J1550-564 
is thought to be powered through the accretion process from a low mass companion 
star. The diversity of the exhibited outburst features depends on the
evolutionary status and the possible irradiative heating of the companion star 
(See, e.g., Tanaka \& Lewin 1995 for a review). XTE J1550-564 was first discovered 
by the all-sky monitor (ASM) on board the \textit{Rossi X-Ray Timing Explorer} 
(RXTE) on 1998 September 7 (MJD 51063) (Smith, 1998) and by the Burst and Transient 
Source Experiment (BATSE) on board \textit{Compton Gamma Ray Observatory} (CGRO; 
Wilson et al. 1998). The Optical (Orosz, Bailyn \&  Jain 1998) and radio 
(Campbell-Wilson et al. 1998) counterparts were detected shortly afterward. 
The optical photometry during this outburst revealed a binary period of 
$1.541\pm0.009$ days (Jain et al. 2001). During the September 1998 outburst, 
a superluminal jet was found to be associated with a massive X-ray flare 
(Hannikainen et al. 2001). Recent observation established that the black hole 
in XTE J1550-564 has a mass of $10.0\pm1.5M\odot$ (Orosz et al. 2002) 
and the companion star is a late-type sub-giant ($G8IV-K4III$). The binary 
inclination angle is $72^{\circ}\pm5^{\circ}$ (Orosz et al. 2002).

In this paper, we revisited the decade old `outburst' especially focusing on the 
timing properties and try to interpret the results in the light of the current
understanding of the accretion flow dynamics around a black hole. 
We thoroughly analize the data of the first three weeks of
the very initial stage of the outburst which includes a rising phase and 
the preliminary declining phase. We clearly observe a very smooth day to day variation of the
QPO frequency in these phases. However, unlike the 2005 outburst of GRO J1655-40
(Chakrabarti et al. 2005; Chakrabarti et al. 2008, hereafter referred to as CDNP08)
where the QPO frequency is seen to steeply rise and disappear 
at the end of the onset phase and to go down monotonically after several months in the
decline phase, QPOs never really disappeared between the rising phase and the declining phase in the
present case. Indeed, there is a continuity in QPO frequency in the rising and decline phases. This 
is very interesting and should throw lights on any model which attempts to explain the evolution of QPOs.
We discuss the possible reason for displaying such a behaviour.

In the literature, a considerable progress has been made in the analysis and interpretation of 
the spectral and timing properties of XTE J1550-564 during its last several outbursts. Soria et al. (2001)
and Wu et al. (2002) showed that while the hard X-ray flux from BATSE 20-200keV 
observation of the 1998 burst reached its maximum after one day, and started to decline in the next 3-4 
days, the soft X-ray flux from RXTE/ASM (2-12keV) continued to rise monotonically for 10 days or so. 
Subsequently, both the hard and the soft X-rays flared after 12 days of the initial hard 
X-ray spike. They also reported a radio flare after $1.8$ day of this flare. This interesting
behaviour lead to the conclusion that both the low and the high angular 
momentum flows could be present simultaneously as in Chakrabarti \& Titarchuk (1995). 
Sobczak et al. (2000a) presented a complete spectral study of XTE J1550-564 
where they presented results of 209 observations spanning 250 days. The whole 
eruption was found to be double peaked, and while the first half
was dominated by a power-law emission, the second half was dominated by the emission
from an accretion disc. Sobczak et al. (2000b) compared the nature of the 1998 outburst of 
XTE J1550-564 with the 1996 outburst of GRO J1655-40 and discussed the similarities
and differences. They find that both exhibited a general increase in QPO frequency with the 
disc flux. QPOs are found to be present only when the power-law component contributes 
more than 20 percent of the 2-20keV flux, thus agreeing with the general perception that
only the Comptonized photons take part in QPOs (Chakrabarti \& Manickam, 2000; hereafter CM00;
Rao et al. 2000). Reilly et al. (2001), using the result from Unconventional Stellar Aspect (USA)
Experiment on board the \textit{Advanced Research and Global Observation Satellite} (ARGOS), showed 
that the centroid frequency of low frequency QPOs during 
the 2000 outburst of XTE J1550-564 tends to rise with increasing USA flux in 1-16keV. They
study the correlations of the hard and soft fluxes and concluded that the observations could
be explained only if the flow has two independent components, one Keplerian and the
other sub-Keplerian. However, none of these papers addressed the 
issue of the evolution of the QPO frequency from the perspective of theoretical understanding.

Although the generic flaring in X-rays are often called the 'outbursts', it possible that all the
outbursts need not involve waves of matter rushing into the black hole in the {\it identical} way.
In CDNP08 it was shown that in GRO J1655-40,  towards the end of the rising
phase of the outburst, the QPO frequency rapidly rose and disappeared altogether. It appeared 
that an oscillating shock wave, which caused QPOs, drifted slowly through the incoming 
flow and eventually disappeared behind the horizon. After several months, in 
the decline phase of the outburst, the oscillatory shock formed close to the 
black hole propagated outward till the rms value of QPO was too weak to be detected. 
If we follow a similar analysis as in CDNP08, we note that in the present case,
the shock never reached close to the horizon before turning back. This gives rise 
to our conjecture that possibly a full-fledged outburst did not take place in this 
object in 1998.

The plan of the paper is the following: In the next section we discuss a 
possible cause of low and intermediate frequency QPOs in accretion discs around 
compact objects, namely, oscillations of shock waves. In \S 3, we present the 
observational results on XTE J1550-564 in detail and show our model fit of the 
QPO frequencies from day to day. From this, we extract the shock parameters. 
In \S 4, we discuss the implications of the present analysis. Finally in \S 5, we
make concluding remarks.

\section{Low and Intermediate frequency QPOs}

A brief description of the physical processes in Low Frequency QPOs (LFQPOs) is in CDNP08 
and we discuss them here only for the sake of completeness.
A satisfactory model of LFQPOs claims that the X-ray oscillation could be due to the
oscillation of the Comptonizing region which is the region between the
centrifugal pressure supported shock and the innermost sonic point
(Chakrabarti, Acharyya \& Molteni, 2004; hereafter CAM04, and references therein).
Perturbations inside a Keplerian disc has also variously been conjectured to be
the cause of low-frequency QPOs (e.g., Trudolyubov, Churazov \& Gilfanov, 1999;
see, Swank 2001 for a review). To our knowledge, no numerical simulations have shown 
that hot perturbing blobs embedded in Keplerian discs can sustain itself beyond a few
dynamical time scale. However, several numerical simulations of accretion flows including the thermal
cooling effects (Molteni, Sponholz \& Chakrabarti, 1996 [hereafter MSC96], CAM04; Okuda et al. 2007) 
or dynamical cooling (through outflows, e.g., Ryu, Chakrabarti \& Molteni, 1997) clearly 
demonstrated that the shocks oscillated with frequencies similar to the 
observed QPO frequencies. The post-shock region does behave
as a Comptonizing cloud for all practical purposes (Chakrabarti \& Titarchuk, 1995).
The shock moves inward with the increase of the cooling rate (MSC96) and it can
propagate when the viscosity is turned on (Chakrabarti \& Molteni, 1995). Using these 
physical results, we have recently explained the way the QPO frequency evolved in the 2005 X-ray outburst 
of GRO J1655-40 (Chakrabarti et al. 2008) quite satisfactorily.

It is to be noted that the shocks are quite common in low-angular momentum (sub-Keplerian) flows and 
they have been extensively studied in the literature (Chakrabarti, 1989;  Nobuta \& Hanawa, 1994; 
Yang \& Kafatos 1994, Lu et al. 1997). There are increasing observational 
evidences both from spectral and timing observations that an accretion process must have a significant 
amount of low angular flow along with the Keplerian flow (Smith et al. 2001, 2002, 2007; Soria et al.
2001; Wu et al. 2002; Reilly et al. 2001). It is thus expected that the other consequences 
of a low angular flow, namely, the shock waves would also be manifested and their signatures 
would also be observed. Thus the past successful explanations of QPO properties with shock
oscillation (CM00; Rao et al. 2000; CAM04) are not far fetched. Indeed, in the case of
1998 outburst of XTE J1550-564, Wu et al. (2002) invoked shocks in the low-angular flow
to explain the observations while Reilly et al. (2001) found the shock oscillations to be most 'natural'
to explain the QPOs for the 2000 outburst. In the shock oscillation model, the
QPO frequency is the inverse of the infall time $t_{infall}\sim  r_s/v \sim  R r_s(r_s-1)^{1/2}$, where,
$R$ is the shock strength (ratio of the post-shock to pre-shock densities), 
$r_s$ is the shock location in units of the Schwarzschild radius $r_g$, and $v$ is the 
radial velocity of the flow in the post-shock region $v \sim 1/R (r_s-1)^{-1/2}$ in units of the velocity of
light (MSC96; CM00; CDNP08). Of course, to trigger the oscillations, the accretion 
rate should be such that the cooling time scale roughly match the infall time scale from the 
post-shock region (MSC96). Thus, the instantaneous QPO frequency $\nu_{QPO}$ (in $s^{-1}$) is expected to be,
$$ 
\nu_{QPO} = \nu_{s0}/t_{infall}= \nu_{s0}/[R r_s (r_s-1)^{1/2}].
\eqno{(1)}
$$
Here, $t_{s0}^{-1}= c/r_g=c^3/2GM$ is the inverse of the light crossing time of the black hole
of mass $M$ in $s^{-1}$ and $c$ is the velocity of light. In a drifting shock scenario,
$r_s=r_s(t)$ is the time-dependent shock location given by,
$$
r_s(t)=r_{s0} \pm v_0 t/r_g.
\eqno{(2)}
$$
Here, $r_{s0}$ is the shock location when $t$ is zero and $v_0$ is the shock velocity
in c.g.s. units. The positive sign in the second term is to be used for an outgoing
shock and the negative sign is to be used for the in-falling shock. 

The shock strength $R$ need not be fixed as it moves towards the black hole. This is because 
general relativistic properties of the horizon will not allow a density
gradient on the horizon. Thus at the most the shock would have a strength $\sim 1$
on the horizon. In case the strength is reduced to $\leqslant 1$ somewhere inside the
disc (which means a sub-sonic to super-sonic transition, which is unstable), 
the shock would rather propagate outward to increase it again. We show below that
this is perhaps what is happening in this object.
                                                                                                      
\section{Observation and Data Analysis}

In this paper, we concentrate on the data of 27 Observational IDs (corresponding to a total of 
20 days) of XTE J1550-564 due to RXTE Proportional Counter Array (PCA; Jahoda et al., 1996). 
We extracted the light curves (LC), the power density spectra (PDS), the energy spectra and the photon counts 
for different energy channels from the best calibrated PCA detector unit, namely, PCU2. 
We used FTOOLS software package version 6.1.1 and XSPEC version 12.3.0. For the timing 
analysis (LC and PDS) we used Science Data of the Normal Mode.

\begin{figure}
\vbox{
\vskip 0.1cm
\centerline{
\includegraphics[width=10.0cm]{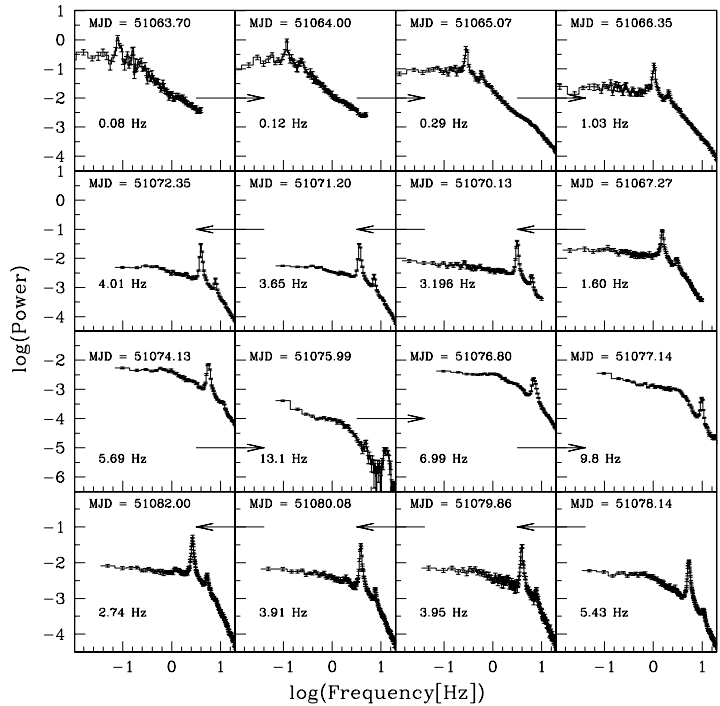}}
\vskip 0.1cm
\caption{Power density spectra during the first few days of the outburst. A
systematic observed shift in the QPO frequency are correlated by a
drifting of the oscillating hot electron cloud in the post-shock
region at a constant velocity. The day of observation and the 
corresponding frequencies are given in the respective boxes.}
}
\end{figure}

\begin{figure}
\vskip -0.2 cm
\centerline{
\vbox{
\vskip -0.2 cm
\includegraphics[width=10.0cm,angle=0]{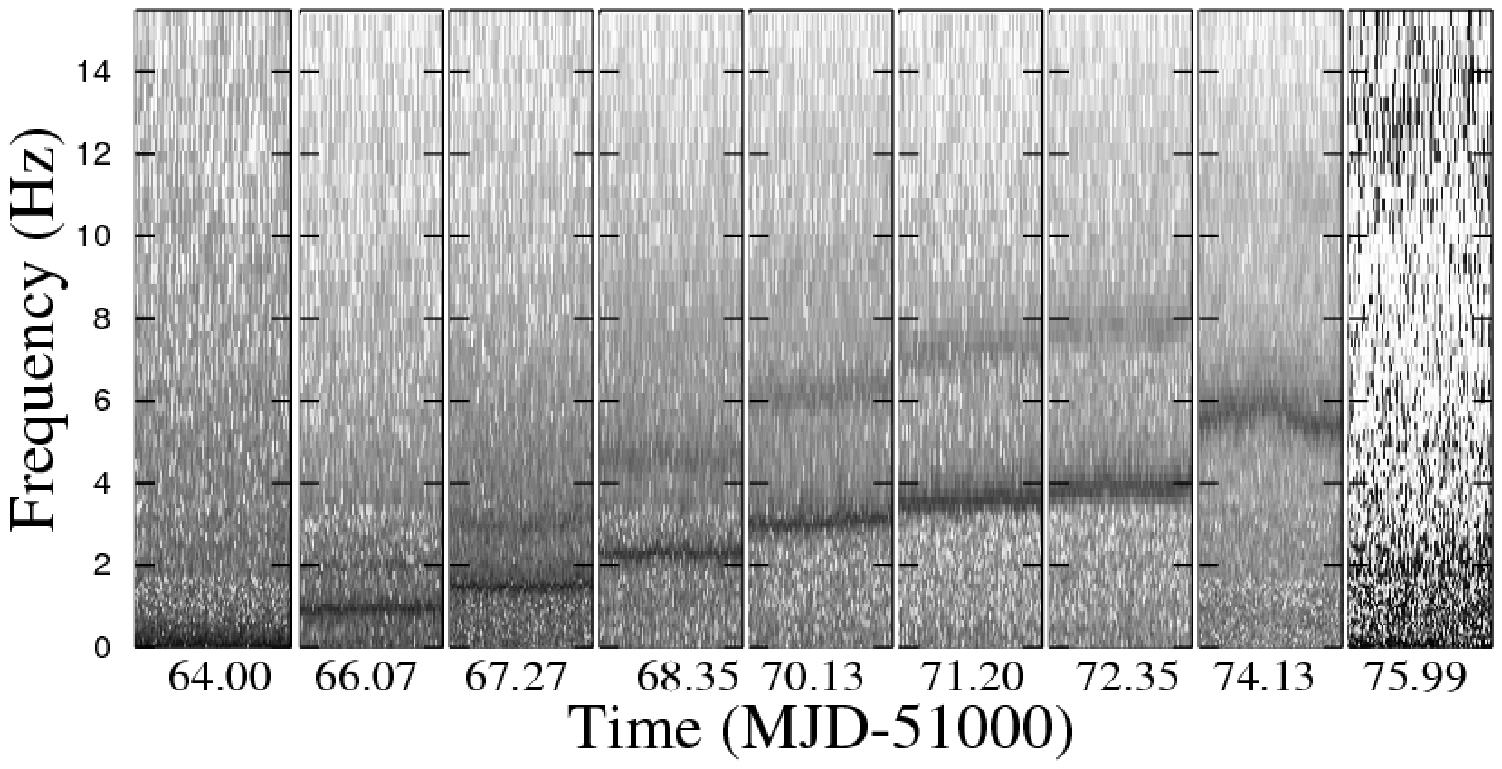}\\
\vskip -0.9 cm
\includegraphics[width=10.0cm,angle=0]{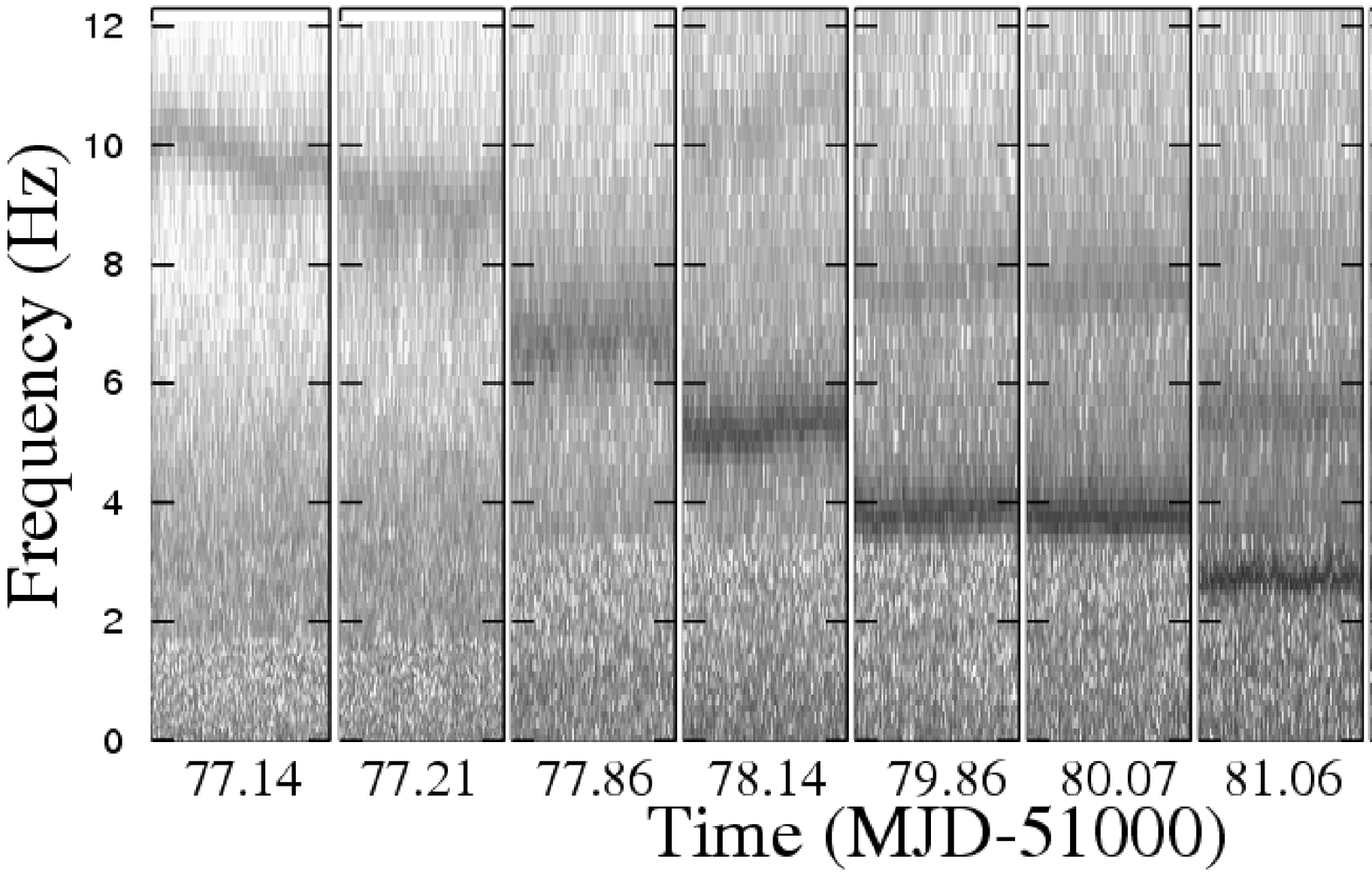}}}
\vskip -0.5 cm
\caption{Dynamic power density spectra of observations (a) in the rising phase and 
 (b) in the declining phase. The vertical direction indicates 
the QPO frequency and the horizontal direction indicates the 'Day'.
Here, MJD 51000='0 Day'. The intensity in the gray scale signifies power. The gray scale
indicates the $\{log(power)\}^2$ and normalised to the range 0-20 except on the $75.99$th
day when the gray scale was normalized to 30-40, to show more contrast.
}

\end{figure}

Figure 1 shows the systematic variation of the power density spectrum covering the first three
weeks of the outburst. QPO frequencies obtained after fitting a Lorentzian are indicated inside
each box. The arrows indicate the direction of increasing time. We clearly observe 
that after reaching the highest value of $\nu_{qpo}=13.1$Hz, the qpo frequency 
started decreasing. Surprisingly, the power itself was very low on this day (see, Fig. 1).
In Figs. 2(a-b), the dynamic power density spectrum for observations (a) in the
rising phase and (b) in the declining phase are shown in the gray scale. At the 
bottom of the panels the start time of each panel is shown after subtracting the MJD values. 
To show better contrast, the square of log (Power) has been plotted with the gray scale range from 
0-20 in all the days of the rising phase and decline phase except for the day 75.99 (MJD=51075.99)
where gray scale range is 30-40.


\begin {figure}
\vbox{
\vskip 1.0cm
\centerline{
\includegraphics[width=8.0cm]{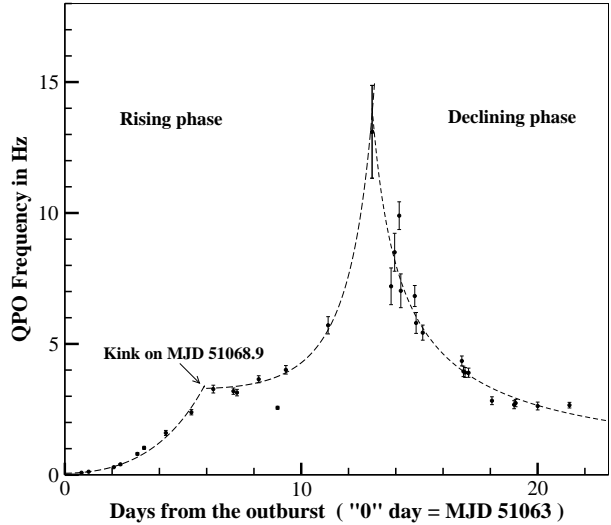}}
\vskip 0.1cm
\caption{Variation of QPO frequency with time (in day) in the rising and decline phases of QPO
since the beginning of the outburst. Error bars are FWHM of fitted Lorentzian curves in the power 
density spectrum. The dotted curves are the solutions from oscillating and propagating shocks. In the 
rising phase, a kink due to the sudden change in shock strength is observed after the $6$th day 
from the beginning of the detection of QPOs. The shocks appears to be drifting at a constant speed 
of $1981$cm/s towards the black hole for best fit. In the declining phase, the shocks 
appears to move away at a constant speed, though there are considerable fluctuations
presumably due to interaction of the outgoing shock and incoming Keplerian disc.
}
}
\end{figure}

In Fig. 3, we show the variation of the QPO frequencies as a function of number of days passed after the 
first detection on MJD 51063 which is chosen to be the zeroth day. The full widths at half maxima 
of the fitted QPOs have been used as the error bars. In the rising phase, $0^{th}$ day starts 
on MJD=51063. As presented in CDNP08, we assume that in the rising phase,
the variation of QPOs are due to slow drifting of the oscillating shocks towards the black hole. In the 
decline phase, the oscillating shock propagates outwards. The dashed curve represents 
our fit with Eqs. (1-2). There appears to be a kink on MJD=51068.9.
The value of reduced ${\chi}^2$ for the fit in the rising phase is $1.74$ up to the 
kink and $0.52$ thereafter. The fit requires that a strong shock ($R=7.0$) is to be launched 
at $r_s=815.0$ which drifts slowly at $v_0=1981$cm s$^{-1}$ towards the black hole.
At the time of the flare on MJD=51075.99 ($12.30$ d), the oscillation frequency was detected at $\nu=13.1$Hz.
At this point, our fit showed that the shock was located at $r \approx 112.4$.
The shock is assumed to be time dependent and  becomes weaker as it propagates 
towards the black hole. This is expected since at the inner boundary, i.e., the horizon,
a density or velocity jump cannot be sustained
and $R \rightarrow 1$ as $r\rightarrow r_g$. For simplicity, we assume
$1/R\rightarrow 1/R_0 + \alpha t_d^2$, where $\alpha$ is a very small number limited by the
time in which the shock disappears ($t_{ds} \sim 12.30$ days): $\alpha > 0.75/t_{ds}^2$. 
Thus in all, we have three free parameters, namely, the initial shock location and velocity and $\alpha$.
In our fit, $\alpha = 0.108$ up to the $6^{th}$ day when the kink was observed and after that shock 
behaviour changed and $\alpha = 0.022$ had to be used. 

The declining phase started almost immediately (within a day) after the rising phase ended on
MJD=51075.99. This behaviour corresponds to the drifting of shocks in the reverse direction.
{\it This is unlike the outburst in GRO J1655-40 (CDNP08) where, the QPO disappears completely in the
rising phase and the decline phase started only after several months.}
The shock was found to drift with time till MJD=51083.00 when $ \nu_{QPO}=2.628$Hz.
It evolves as $ \nu_{QPO} \sim t_d^{-0.8}$ and since $ \nu_{QPO} \sim r_s^{-2/3}$ (Eq. 1), 
$r_s \sim t_d^{1.2}$. Thus the shock steadily recedes away from the black hole
with almost constant velocity ($v_s \sim t_d^{0.2}$). The value of Reduced $ {\chi}^2$ for this fit (with three degrees of freedom) 
is $1.09$ and thus the fit is satisfactory. After that the oscillation frequency varied between 
$3$-$6$Hz for about $26$ days till MJD=51109.74 (see also, Sobczak et al. 2000a). This shows that 
The shock started stalling exactly as in GRO J1655-40 (CDNP08). We believe that this behaviour is
due to the interaction of the receding shocks with incoming Keplerian flow, which continued to 
accrete due to longer viscous timescale. Beyond MJD 51,109.7 the Keplerian 
disc is drained in a time scale of $\sim 11$ days (Wu et al. 2002) while the 
propagating shock in the sub-Keplerian flow, could be too weak to have any
observational effect. Thus no regular QPO was seen in this phase. This is in contrast 
with GRO 1655-40, where the shock receded at a constant acceleration towards the
very end of the decline phase.

\begin{figure}
\vbox{
\vskip 0.1cm
\centerline{
\includegraphics[width=8cm]{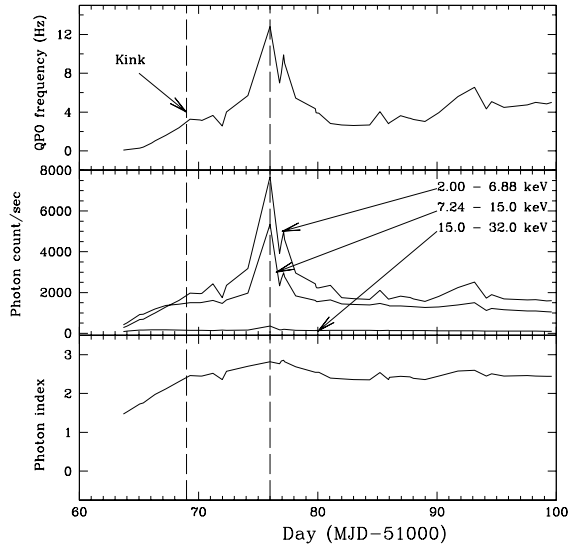}}
\vskip 0.1cm
\caption{A comparative diagram of QPO frequency, Photon count for different energy bands 
and the photon index with day. Here we choose MJD 51000= '0 day'. In the first few
days, the QPO frequency rises sharply with intensity of the soft X-rays. 
In the declining phase, the QPO frequency goes down rapidly in the
first few days and then started to fluctuate appearing only sporadically, in the 
$3-6$Hz range in the subsequent $26$. The spectrum was marginally soft during this phase.}
}
\end{figure}

In Fig. 4, we plot the variation of QPO frequency (upper panel), the fluxes of 
the hard, the intermediate and the soft X-rays (middle panel) and the power-law 
photon index (lower panel). Since the shock oscillation causes QPOs in this model,
both the pre-shock flow responsible for very soft photons
and the post-shock flows responsible for the Compotonized photons, become hotter and luminous
as the shock approaches the black hole. The evolution of the photon spectral index 
shows that generally the spectrum became softer as the rising phase proceeded
and became harder only marginally in the declining phase. In fact, the spectral index 
changed very little during the final twenty days (see, Fig. 4). The 
photon count is neither low (as in the beginning) nor high (as in the flare)
and the index is marginally soft. Thus the system is in an intermediate state.

\begin{figure}
\vbox{
\vskip 0.1cm
\centerline{
\includegraphics[width=10cm]{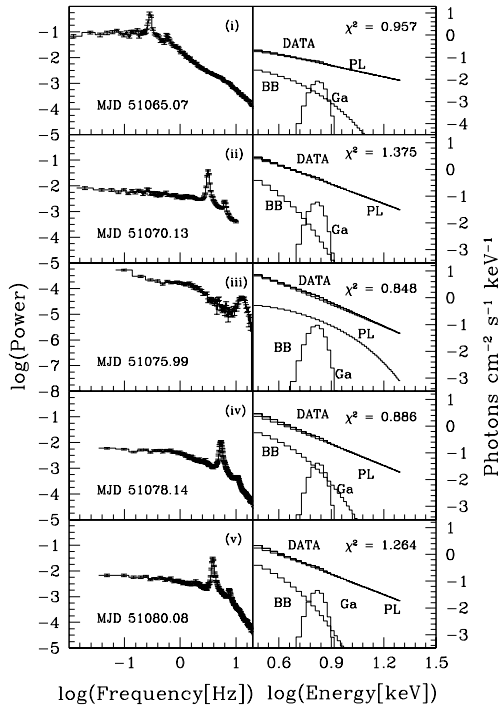}}
\vskip 0.1cm
\caption{Power density spectrum (left panel) and energy spectral index (right panel) of the rising phase (i-iii)
 and the declining phase (iv-v). The QPO frequencies are (i) 0.29Hz (0.957), (ii) 3.2Hz (1.375) (iii) 13.1Hz 
(0.848) (iv) 5.43Hz (0.886), (v) 3.90Hz (1.264) respectively. Numbers next to frequencies are the 
reduced $\chi^{2}$ values. The component used for fitting the energy spectra are marked: 'DATA' for total 
fit, 'BB' for black body, 'Ga' for Gaussian, 'PL' for a power-law component without a cut-off.}
}
\end{figure}

The physical picture during these initial days 
of the 1998 episode becomes clearer when we analize both the timing 
and the spectral properties. We plot the results for a few selected days 
in Fig. 5 (panels i to v). The left panels show the power density spectra
and the right panels show the energy spectra. 
The Observation IDs (marked in each panel) in the rising phase are
30188-06-01-00 ($2^{nd}$ day), 30188-06-07-00
($7^{th}$ day) and 30191-01-02-00 ($12^{th}$ day) and in the decline phase are
30191-01-08-00 ($15^{th}$ day) and 30191-01-10-00 ($17^{th}$ day) respectively. 
In the rising phase, the spectrum becomes softer as the shock moves in and the QPO frequency
rises. The softening of the spectrum with the increase in the QPO frequency
has been reported before (e.g., Chakrabarti et al. 2005;
ST06; Shaposhnikov et al., 2007; CDNP08). 
MSC96 and CAM04 explicitly showed that an increase in the cooling rate reduces 
the shock location and thus increases the QPO frequency. We observe 
that the black body (BB) and the Gaussian (Ga) component
from the Keplerian disc are strengthened by factor of ten in between 
the panel (i) and panel (iii). A similar behaviour was also seen by ST06
using the data of Cyg X-1. A kink on day $6^{th}$ in the rising phase 
is peculiar in the sense that the rising trend of the power-law photon
index is decreased abruptly (lower panel of Fig. 4) and the rising trend 
of QPO frequency (Fig. 3) also changed abruptly. It is possible that 
there are multiple shock waves in the flow, both in the rising phase 
and in the decline phase.

\section{Discussions}
 
\begin{figure}
\vbox{
\vskip 0.1cm
\centerline{
\includegraphics[width=9cm]{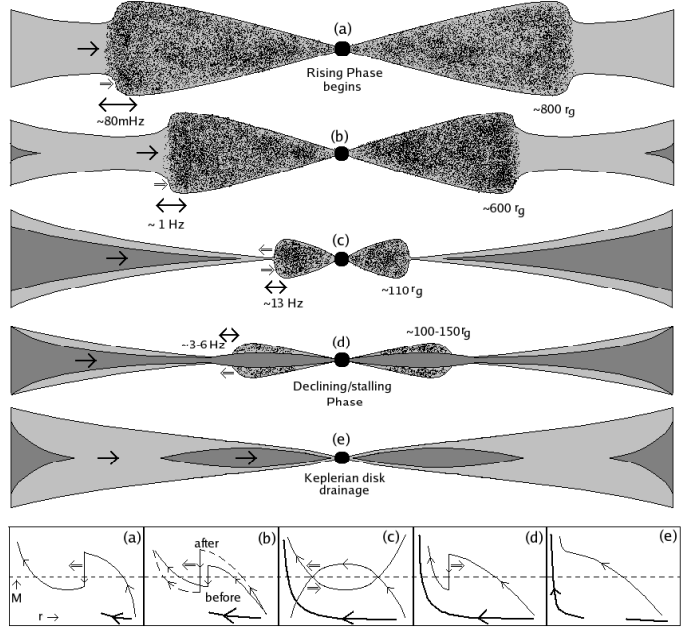}}
\vskip 0.1cm
\caption{A physical picture which emerges out of the spectral and temporal analysis of the 1998 
outburst of XTE J1550-564: (a) beginning of the rising phase when QPO was first detected , 
(b) middle of the rising phase, where the Keplerian disc also started to move in, (c) end of the rising
phase and the beginning of the declining phase, (d) stalling of the outgoing shockwave due to
interaction with the incoming Keplerian disc and finally (e) the draining of the part of the
Keplerian component which ended the first phase of outburst. The lowermost panel gives a cartoon diagram of the
Mach numbers of the Keplerian and the sub-Keplerian components (Chakrabarti 1989).}
}

\end{figure}

Analysis of the 1998 outburst gives a clearer physical picture of what is happening 
in XTE J1550-564. Fig. 6 shows the plausible accretion scenario in different 
phases (a-e) of the outburst. The lowest panel shows the corresponding cartoon diagrams of the
Mach number (M) vs. radial distance (r) of the Keplerian (thick curve) and the sub-Keplerian (thin curve)
flow (Chakrabarti, 1989). In panel (a), at the beginning of the rising phase, the 
flow is dominated by a sub-Keplerian flow with a shock at around $\sim 800 r_s$. The
shock propagates inward with a constant velocity and  at the same time, the Keplerian
disc moves in viscous time scale. As a result the spectrum becomes softer and the
QPO frequency also rises [panel (b)]. The shock strength decreases and becomes $\sim 1$
at around $r_s \sim 500 r_s$ when the QPO frequency was around $3.3$Hz. At this stage
($\sim 6$ days after the outburst began) either the
same shock strengthened or a new wave of sub-Keplerian matter moved in and the QPO
behaviour was changed distinctly (Fig. 3). Subsequently [panel (c)], the shock becomes
weaker and stopped while a reverse shock formed and propagated outward to initiate 
the declining phase. The blackbody component is strongest in this case. The reverse shock
propagated outward at almost constant velocity as QPO frequency also started to decrease [panel (d)].
The supply of Keplerian flow also ceased and the flow broke into two parts, the inner part is moving in
while the other part stayed back. The segment moving in is eventually drained in viscous timescale
($\sim 11$ days; Wu et al. 2002) [panel (e)].

\section{Concluding remarks}                                                                                                         
In this paper, we studied the evolution of the QPO frequencies in the 1998 outburst of the 
black hole candidate XTE J1550-564. During the rising phase, the QPO frequency increases 
very slowly first few days and then very rapidly. The spectrum also 
became softer very closely following the QPO behaviour.
Immediately after reaching a frequency of $\sim 13.1$Hz, the decline phase started which is 
characterized by softening of the spectrum. The QPO frequency started to decrease also immediately.
The spectral and temporal analysis suggest that the smooth variation of the QPO
frequency is due to the movement of the Comptonizing region itself. A satisfactory
model of the low frequency QPOs which claims that the oscillation of the shock wave is 
responsible for QPOs can also explain the daily variation provided the shock propagates towards the
black hole at a velocity of about $20$m/s. There was a kink in the frequency variation 
in the rising phase, indicating a possibility of sudden strengthening of the same shock wave or 
the presence of another incoming shock.

A comparison of the 2005 outburst in another transient source GRO J1655-40 (CDNP08) suggests 
a number of similarities and dissimilarities. For instance, a similar shock velocity was also necessary 
there to explain the QPO frequency variation. 
In both the objects the spectrum was marginally soft, i.e., the photon index is just above $2$ after the
decline phases. This phase appears turbulent, with sporadic 
QPOs, as though the outward movement of the shockwave is confronting the still incoming Keplerian 
flow which moves in with a much longer viscous timescale. However, there are several major 
differences: (a) the nature of the rising phase differs qualitatively. In GRO J1655-40, 
the shock wave moved closer to the black hole and ultimately disappeared behind the horizon. 
In the present case, the best fit curve required that shock strength to go down to less than unity
much before reaching the horizon. Thus the shock wave never reaches the horizon. This could be 
either because of a reverse shock in the sub-Keplerian flow which overpowered this incoming wave
or the shock started to propagate in the reverse direction due to pressure imbalance.
The outward propagating shock in GRO 1655-40 started only after several month, indicating that
matter supply was prolonged while in the present situation, the supply was erratic causing
the receding shock to form immediately after the flare.
In GRO J1655-40, the shock was stalling in the decline phase only for $3.5$ days, 
while in the present case, the shock was receding with almost 
constant velocity ($v_s \sim t^{0.2}$) in the first week and then stalled for the next four weeks. After that
QPO also disappeared.

In majority of the black hole candidates, existence of two component flows has been proven 
beyond any reasonable doubt. Even in this object, workers have always invoked two components
to explain the spectral variations. The sub-Keplerian flow is found to be responsible 
for supplying hot electrons, which produce the power-law component through inverse Compton
process. However a major consequence of a sub-Keplerian flow, namely to produce 
an oscillating shock was overlooked. This oscillation is due to the tug-of-war between the
cooling process which tends to collapse the shock and centrifugal barrier which 
tends to push the shock outward (MSC06). In this paper, we show that the 
oscillating shock is probably responsible for QPOs which can propagate forward or
backward depending on the extent to which the Rankine-Hugoniot condition is broken at the shock.
In the rising phase, the excess cooling in the post-shock flow (which is the CENtrifugal 
pressure dominated BOundary Layer or CENBOL) causes a steady drift of the shock
towards the black hole, while in the decline phase the recession of the Keplerian disc
causes the reduction of pressure in the pre-shock flow and the shock can propagate outward.
What is important is that we now have a consistent picture of how low and intermediate frequency 
QPOs are generated and are evolved in accretion flows around black hole candidates
during the outburst. To our knowledge there is no competing model of low and intermediate frequency
QPOs which explain the frequencies as well their variation with decent
reduced $\chi^2$ of our fits. On the other hand, we are puzzled as to why in more than one black hole
candidates, the propagation speed of the shock is $\sim 20$m/s. 

B.G. Dutta acknowledges the support of "Teacher Fellowship" award under the Faculty Improvement 
Programme (F.I.P) scheme of U.G.C.  and S.K.C. and P.S. Pal acknowledges the support of ISRO RESPOND project.
We thank Dr. A.Nandi for helpful discussions on the data analysis.

{}

\end{document}